\documentclass[12pt]{article}%
\usepackage[utf8]{inputenc}%
\usepackage[english]{babel}%
\usepackage[dvips]{graphicx}%
\usepackage{hhline}%
\usepackage{longtable}%
\begin{document}

\begin{center}
{\bf \Large  Anatomy of Ursa Majoris}

\bigskip

I.\,D.\,Karachentsev, O.\,G.\,Nasonova, H.\,M.\,Courtois

\bigskip

Accepted MNRAS 2012 November 26 ; in original form 2012 September 7

    \bigskip                       {\bf  Abstract}
                           
\end{center}

A nearby friable cloud in Ursa Majoris contains 270 galaxies with radial
velocities $500 < V_{LG} < 1500$~km~s$^{-1}$ inside the area of
RA$=[11^h\hspace{-0.4em}.\,0, 13^h\hspace{-0.4em}.\,0]$ and
DEC$=[+40^{\circ}, +60^{\circ}]$. At present, 97 galaxies of them have
individual distance estimates. We use these data to clarify the structure and
kinematics of the UMa complex.

According to Makarov \& Karachentsev (2011), most of the UMa galaxies belong to
seven bound groups, which have the following median parameters: velocity
dispersion of 58~km~s$^{-1}$, harmonic projected radius of 300~kpc, virial mass
of $2\cdot 10^{12} M_\odot$, and virial-mass-to-$K$-band-luminosity of
$27 M_\odot/L_\odot$. Almost a half of the UMa cloud population are gas-rich
dwarfs (Ir, Im, BCD) with active star formation seen in the GALEX UV-survey. The
UMa groups reside within 15--19~Mpc from us, being just at the same distance as
Virgo cluster. The total virial mass of the UMa groups is $4\cdot 10^{13} M_\odot$,
yielding the average density of dark matter in the UMa cloud
to be $\Omega_m = 0.08$, i.e. a factor three lower than the cosmic average.
This is despite the fact that the UMa cloud resides in a region of the Universe
that is an apparent overdensity. A possible explanation for this is that most
mass in the Universe lies in the empty space between clusters.
Herewith, the mean distances and velocities of the UMa groups follow
nearly undisturbed Hubble flow without a sign of the ``Z-wave'' effect caused by infall
toward a massive attractor. This constrains the total amount of
dark matter between the UMa groups within the cloud volume.

\section{Introduction}

The Virgo cluster which is the heart of the Local supercluster of galaxies 
locates 17  Mpc away from us. It is sided on the north and on the south with
diffuse filamentary structures which outline, together with the Virgo cluster,
the Local supercluster plane. The Coma~I complex of galaxies is situated
on the north side, just behind the zero velocity surface of the Virgo cluster, 
$R_0=6.8$~Mpc or 23$^{\circ}$. This cloud consisting of several virialized
groups shows fast non-Hubble motions with amplitude $\sim700$~km~s$^{-1}$. Such
motions can be caused by the expansion of the large cosmic void between the Coma
and Virgo clusters, or by the presence of a massive Dark Attractor located in
the Coma~I region $\sim15$~Mpc away from us (Karachentsev et al. 2011).

There is another cloud of bright galaxies further north from the Virgo cluster,
also located in the Supergalactic plane, Ursa Majoris. It was used by Tully et
al. (1996) and Tully \& Courtois (2012) for calibrating Tully-Fisher (1977) relation
(=TF). Unlike the ``hot'' Coma~I region, the ``cold'' cloud Ursa Majoris
(=UMa) has a quite low value of radial velocities dispersion,
$\sim150$~km~s$^{-1}$. According to Tully et al. (1996), the UMa cloud includes
galaxies with radial velocities relative to the centroid of the Local Group
$700<V_{LG}<1210$~km~s$^{-1}$ in the circle of radius $7.5^{\circ}$ and with
centre coordinates $11^h56.9^m+49^{\circ}22^{\prime}$. Galaxies of the late
morphological types with high HI content are the prevailing population of the
UMa complex (Trentham et al. 2001), what makes them an easy-to-use tool for calibrating TF relation.
According to Tully's catalogue (1988), the UMa (or ``12--1'') cloud accounts in
total 79 galaxies with radial velocities lying in the indicated range. Over the
last years, the number of galaxies in this region with appropriate radial
velocities has grown significantly, essentially due to the Sloan Digital Sky
Survey (Abazajian et al. 2009). Recently Wolfinger et al. (2012) carried out 
a special HI survey of UMa to find new gas-rich dwarf galaxies.
Trebling of the number of galaxies with measured
radial velocities in the UMa cloud provides a good base to reconsider structure
and kinematics of this neighbouring galaxy complex with a fresh perspective.

\section{Structure of the UMa cloud}
Compiling a list of possible UMa cloud members we have assumed some more
mild conditions on galaxy coordinates and velocities:
$$11.0^h<RA<13.0^h, \,\, +40^{\circ}<DEC<+60^{\circ},\,\,
+500<V_{LG}<1500$$~km~s$^{-1}$
than those adopted by Tully et al. (1996). The list of 270 galaxies satisfying
these criteria presented in the Table~1. The columns of the table contain:
(1) galaxy name or its number in the known catalogues; the
coordinate nomenclature for SDSS and 2MASX galaxies was omitted; (2) equatorial
coordinates for 2000.0 epoch; (3) radial velocity (km~s$^{-1}$) in the Local
Group frame with the apex adopted in NED (http://ned.ipac.caltech.edu); (4)
morphological type according to the digital de Vaucouleurs scale; since the
considered region is located entirely in the SDSS zone we determined type T
independently from LEDA (http://leda.univ-lyon1.fr) and found considerable
discrepancies in several cases; (5) integral apparent magnitude in the $K_s$
band; the data on $K_s$ for bright galaxies were retrieved from 2MASS Survey
(Jarrett et al. 2000), for more faint galaxies $K_s$ magnitudes were estimated
by apparent B magnitude and average colour index $<B-K>$ for every type T in the
manner described by Jarrett et al. (2000); (6) name of the brightest galaxy in
the MK group (Makarov \& Karachentsev, 2011) which the considered galaxy belongs
to; (7,8) distance modulus and distance D in  Mpc; we used distance moduli
estimates from NED compilation as the main source giving preference to recent
publications (Tully et al. 2009, Springob et al. 2009).  Among 97 galaxies with
$(m-M)$ estimates 4 objects have distances measured by Cepheids and Supernovae,
9 galaxies of early types have distance estimates by surface brightness
fluctuations (Tonry et al. 2001), and for the rest majority their distances were
determined by Tully-Fisher method. The authors have estimated distance moduli
$(m-M)$ for several galaxies with known HI line widths $W_{50}$ and appropriate
inclinations from the Tully-Fisher relation using parameters proposed by Tully
et al. (2009).

\begin{figure}[p]
\begin{center}
\includegraphics[height=0.6\textwidth,keepaspectratio,angle=270]{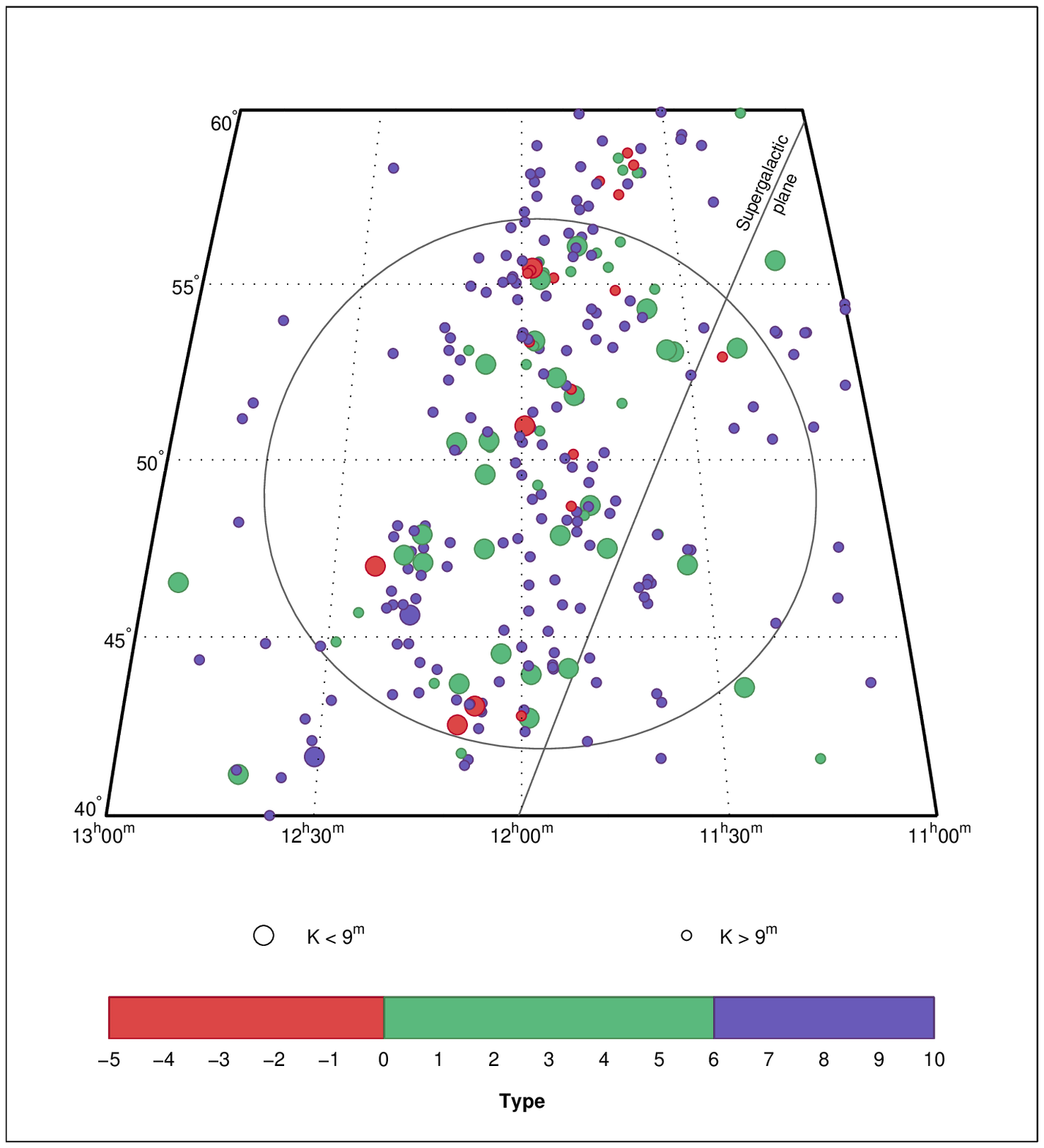}
\includegraphics[height=0.6\textwidth,keepaspectratio,angle=270]{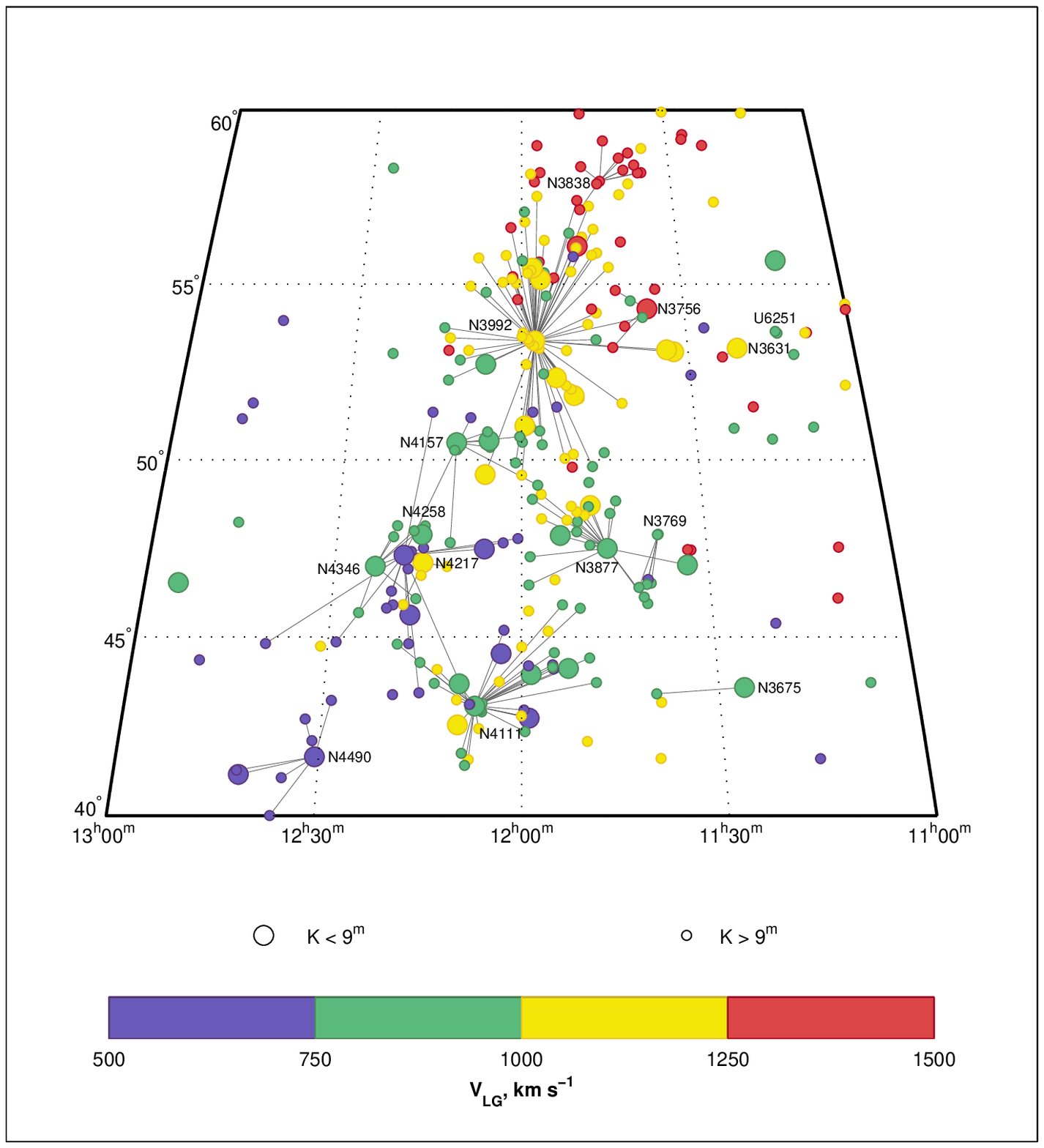}
\vbox{\vspace{0.6cm}
\parbox{1\textwidth}{\footnotesize{}%
Fig.$\;$1. The Ursa Majoris complex of galaxy groups in equatorial coordinates.
{\itshape\bf{}Top:} early type, intermediate type and late type galaxies indicated
by different colours. Bright ($K < 9^m$) galaxies are shown by larger circles.
{\itshape\bf{}Bottom:} the same field with indication of radial velocities of the
galaxies and their membership in different groups.}}
\end{center}
\end{figure}

The distribution of 270 galaxies in the UMa region in equatorial coordinates is
presented in the left panel of Figure~1. Objects of different morphological
types are denoted by circles in different colours. Small circles mark faint
galaxies with apparent magnitudes $K>9.0$. The solid line under the hour angle
$\sim2^h$ corresponds to the Local supergalactic equator. The big circle of
radius $7.5^{\circ}$ shows the region covered by UMa cloud according to Tully et
al. (1996).

As one can see from this diagram, the galaxy complex in UMa does not have any
distinct concentration towards the single centre. The distribution of galaxies
in this region is patchy and elongated along the line of supergalactic equator.
Thus, the UMa cloud seems to be explicitly unrelaxed structure which does not
correspond to galaxy cluster definition. According to the Local supercluster
galaxy groups catalogue (Makarov \& Karachentsev 2011) the UMa cloud contains 7
groups with close values of radial velocities dominated by the galaxies:
NGC\,3769, NGC\,3877, NGC\,3992, NGC\,4111, NGC\,4157, NGC\,4217 and NGC\,4346.
These groups include 159 galaxies or 59\%  from its total number in this region.
Another 45 galaxies (i.e. 17\%) belong to four neighbouring groups (centred
around NGC\,3838, NGC\,4151, NGC\,4258 and NGC\,4490) forming background or
foreground relative to the UMa complex itself. The rest quarter of galaxies
appears as single objects and members of several pairs situated mainly on the
outskirts of the complex. 

Distribution of UMa galaxies in the same area and scale is represented in
the right panel of Figure~1 where group members are linked with dominating
galaxies by straight lines. Four ranges of radial velocities are marked by
different colours. The average radial velocity of galaxies tends subtly to
increase northward along the supergalactic plane. The spiral galaxies NGC\,4258
($K= 5.46^m$) 7.8~Mpc away from us and NGC\,3992 ($K= 6.93^m$) 22.9~Mpc away are
among the brightest objects in the considered region. The first one is
associated with 17 satellites, and this group belongs to UMa foreground
according to Tully et al. (1996). NGC\,3992 galaxy dominates dynamically the
group consisting of 74 members and concentrating about a half of the total
luminosity and mass of the UMa complex.

The group of galaxies around NGC\,4111 ($K= 7.55^m$, $D=15.0$~Mpc) is notable
for some higher content of early types galaxies. Among its seven bright members
with $K<9.0^m$, four belong to S0 and Sa types, which possibly indicates the
advanced evolutionary status of this group.

\section{Hubble flow in the UMa complex}

The Hubble diagram $V_{LG}$ vs. $D$ for 97 galaxies in the UMa region is
presented in the upper panel of Figure~2. Only two groups, NGC\,3992 and
NGC\,4111, have more than 10 galaxies with measured distances(26 and 13
respectively). The members of these groups are shown as squares and circles
linked with dominating galaxies by straight lines. The members of five other
groups of the UMa complex are marked by triangles while the field galaxies are
presented as crosses. The straight line corresponds to the Hubble parameter
value $H_0=73$~km~s$^{-1}$~Mpc$^{-1}$.

\begin{figure}[p]
\begin{center}
\includegraphics[height=\textwidth,keepaspectratio,angle=270]{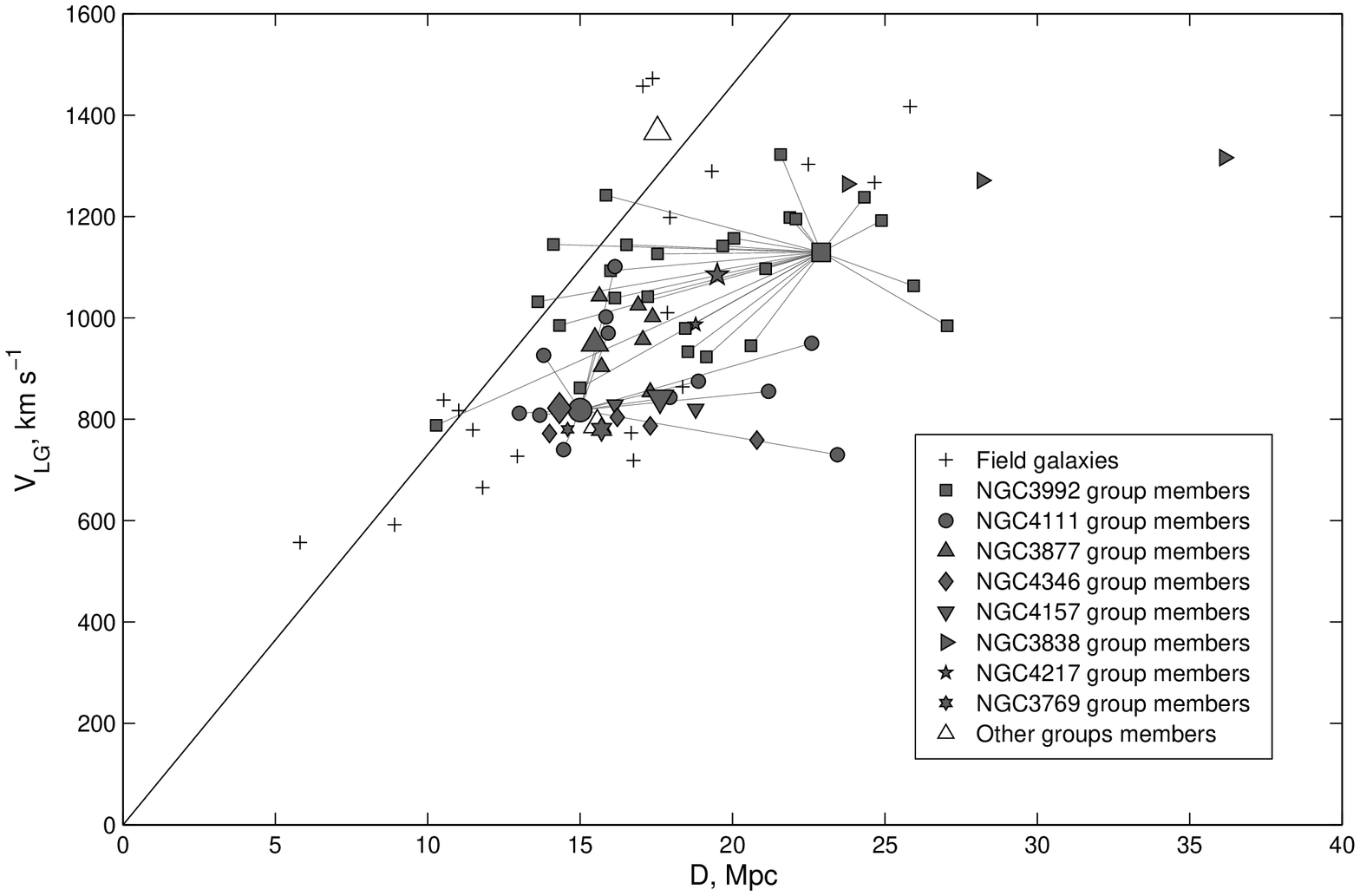}
\includegraphics[height=\textwidth,keepaspectratio,angle=270]{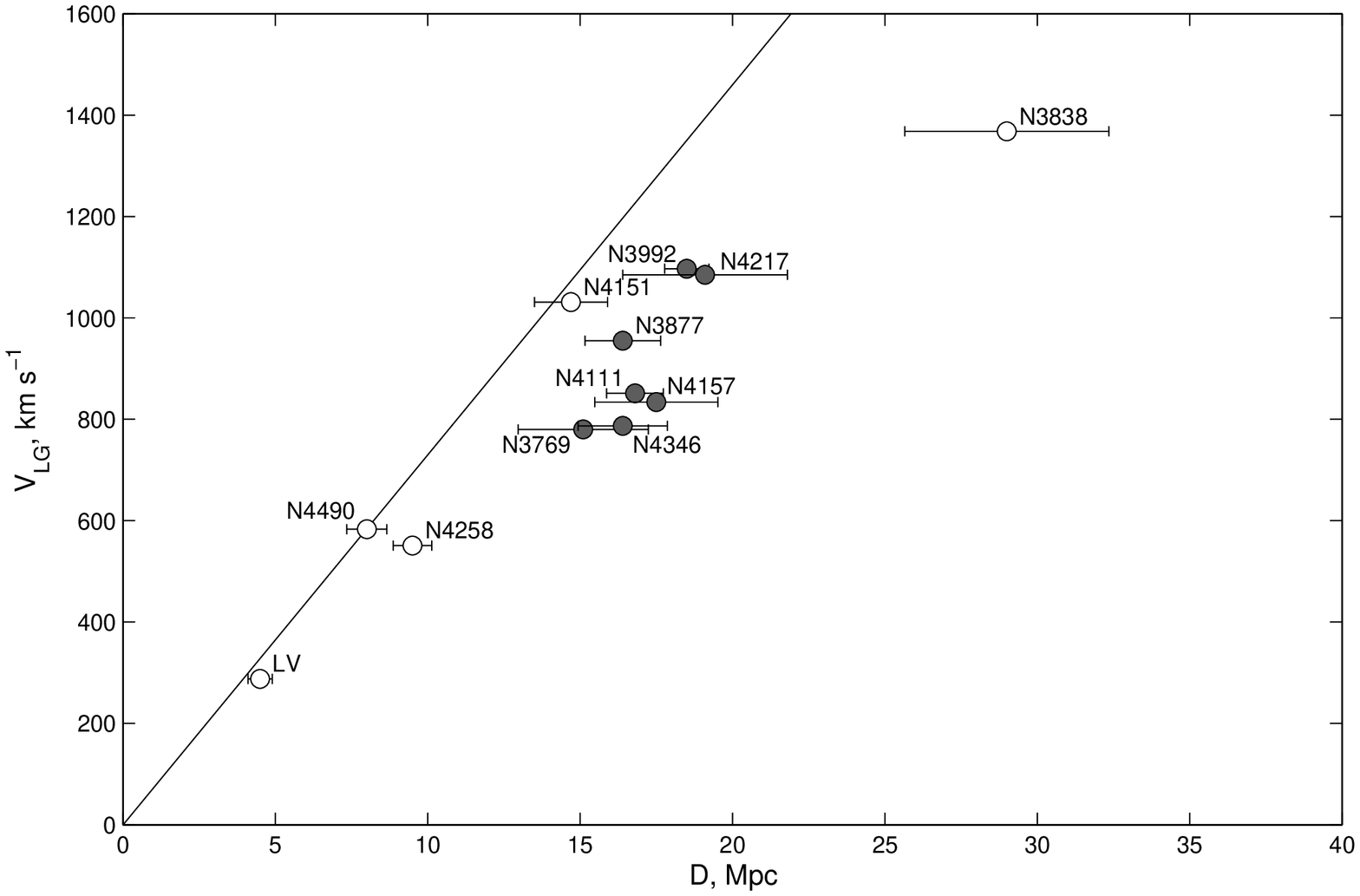}
\vbox{\vspace{0.6cm}
\parbox{1\textwidth}{\footnotesize{}%
Fig.$\;$2. The velocity-distance relation for galaxies in the Ursa Majoris area.
The undisturbed linear Hubble flow with $H_0 = 73$~km~s$^{-1}$~Mpc$^{-1}$ is
shown by the straight line. {\itshape{}Top:} individual galaxies.
{\itshape{}Bottom:} centres of the groups.}}
\end{center}
\end{figure}

As one can see from this diagram, the field galaxies follow generally the Hubble
relation with the parameter $H_0\sim60$~km~s$^{-1}$~Mpc$^{-1}$ and the scatter
of about 20\%, which corresponds to the typical error for Tully-Fisher distance
modulus $\sigma(m-M)\simeq 0.4^m$. Unlike the field galaxies, the members of
NGC\,3992 and NGC\,4111 groups do not show any visible correlation between
velocities and distances as expected for virialized systems. It should be noted,
however, that lenticular galaxies NGC\,3990, NGC\,3998 and NGC\,4026 in the
NGC\,3992 group have $\sim4$~Mpc less distances than an average one for all
other group members. This enigmatic circumstance was mentioned by Tully \&
Courtois (2012). As one can see in the SDSS colour images, some S0 galaxies in
the UMa region, NGC\,3928 for example, have emission and dust patterns which
should affect systematically distance measurements based on the surface
brightness fluctuations method.

The bottom panel of Figure~2 represents the Hubble diagram for centres of the
groups. Every group is depicted as horizontal segment while its length indicates
mean distance error for the group members. There are 11 Local volume galaxies
situated in front of the UMa complex with radial velocities 
$V_{LG}<400$~km~s$^{-1}$ and distances $D<6$~Mpc measured from the luminosity of
the tip of the red giant branch. Their average values for $V_{LG}$ and $D$ are
also shown in bottom panel as ``LV''.

As seen from these data, the foreground objects: LV, NGC\,4490, NGC\,4258
and the neighbour from South group NGC\,4151 follow Hubble relation quite well
with the standard value $H_0=73$~km~s$^{-1}$~Mpc$^{-1}$. However, all the
seven UMa groups have a significant shift to the right going clearly beyond
distance errors. As a whole for seven UMa groups, the mean peculiar velocity
relative to $H_0\sim73$~km~s$^{-1}$~Mpc$^{-1}$ line is
$<V_{pec}>=-337\pm28$~km~s$^{-1}$. Characterising each group by individual value
$H_i=<V_{LG}>/<D>$ brings evidence that all UMa groups
satisfy the condition $H_i=[48-59]$~km~s$^{-1}$~Mpc$^{-1}$ with an average value
$<H_i>=(53.3\pm1.8)$~km~s$^{-1}$~Mpc$^{-1}$. The NGC\,3838 galaxy group lying
behind the complex near its northern boundary yields also some less value
$H_i=47$~km~s$^{-1}$~Mpc$^{-1}$. Kinematic situation in the discussed region
could be described by reference to the ``domain'' concept: the Local group and
the nearby groups (M\,81, Cen~A, NGC\,253, etc.) through NGC\,4490 and NGC\,4258
groups form the ``Local Domain'' or the ``Local Sheet'' characterized by low
value of internal velocities dispersion while seven UMa groups constitute the
neighbouring ``UMa Domain''. Both domains converge just like tectonic plates, that
have mutual approaching velocity of $\sim300$~km~s$^{-1}$~Mpc$^{-1}$. A similar
idea was already developed by Tully et al. (2008) to explain the converging
movement of the ``Local Sheet'' and the Leo cloud.

\section{Dynamical parameters of the UMa domain}

The principal characteristics of the seven UMa groups from Makarov \&
Karachentsev (2011) are represented in Table~2. The table columns contain: (1,
2) name of the dominating galaxy and its equatorial coordinates; (3) mean radial
velocity of the group has been used to determine its Hubble distance assuming
$H_0=73$~km~s$^{-1}$~Mpc$^{-1}$; (4) number of group members with measured
radial velocities; (5) morphological type of the dominating galaxy; (6) radial
velocity dispersion in ~km~s$^{-1}$; (7) mean harmonic radius in kpc; (8)
integral luminosity in the $K_s$ band in solar luminosity units; (9, 10) virial
(projected) mass of the group and virial mass-to-total luminosity ratio in solar
units; (11, 12) mean distance modulus via individual group members and its
dispersion; (13) linear distance in  Mpc corresponding to the mean modulus; (14)
number of group members with distance estimates; (15) radius of the zero
velocity surface (in  Mpc) which is expressed in terms of the total mass of the
group as

$$\log(R_0/ Mpc)=1/3[\log(M/M_{\odot}-12.35]$$
assuming the standard cosmological model parameters
$H_0=73$~km~s$^{-1}$~Mpc$^{-1}$ and $\Omega_{\lambda}=0.76$ (Nasonova et al.
2011).

The second line for each group specifies radial velocities dispersion corrected
for the velocity errors, together with size, luminosity and mass of the group
assuming the mean distance value from column (13) and taking into account radial
velocity measurement errors. The lower part of the Table~2 contains similar data
for four groups neighbouring upon UMa and Coma~I complexes of galaxies.

Seven groups constituting the UMa complex are characterised by the following
median parameters: radial velocity dispersion of 58~km~s$^{-1}$, harmonic radius
of 300~kpc, integral $K$-band luminosity of $15.5\cdot10^{10}L_{\odot}$ and
unbiased virial mass estimate of $2.0\cdot10^{12}M_{\odot}$ coinciding by chance
with the Local group mass estimate. The virial mass-to-total $K$-luminosity
ratio for these groups corrected for velocity errors lays in the range [10\,--\,38]
$M_{\odot}/L_{\odot}$. As a whole for UMa complex, the total luminosity of 7
group members amounts $1.51\cdot10^{12}L_{\odot}$. A small increment to it (8\%)
is added by single galaxies and several pair members distributed among the
groups. The sum of unbiased virial mass estimates for seven groups,
$4.3\cdot10^{13}M_{\odot}$, corresponds to the mass-to-luminosity ratio for the
complex $(M/L_K)=28 M_{\odot}/L_{\odot}$.

Since the relation $M_*/L_K\simeq 1\cdot M_{\odot}/L_{\odot}$ fulfils for stars
in general (Bell et al. 2003), it can be presumed that dark (virial) mass of the
UMa cloud is 27 times more than its luminous (star) mass. This value increases
if there is some additional quantity of dark matter distributed among the groups
of the complex and inappreciable by virial method. Though dark matter dominates
in the UMa complex, its quantity there is not somewhat extraordinary. As it was
noted by Makarov \& Karachentsev (2011), the average global matter density in
standard $\lambda$CDM model, $\Omega_m=0.28$, corresponds to some average value
$(M/L_K)=97 M_{\odot}/L_{\odot}$. Then, the obtained ratio
{28$M_{\odot}/L_{\odot}$ for UMa cloud is expressed in terms of mean densities
as $\Omega_m =0.08$, i.e. considerably lower than the global density. As it
was pointed out by Karachentsev (2012), there is a disagreement between the
mean local estimate of matter density, $\Omega_{m,loc}=0.08\pm0.02$
within a sphere of radius $\sim50$~Mpc and the
global cosmic value $\Omega_{m,glob}=0.28\pm0.03$. The case of UMa complex
provides yet more evidence for the existance of ``Missing Dark Matter'' problem.

Most distances to the UMa domain galaxies are determined through Tully-Fisher
relation with a typical error of $\sim20$\% or $\sigma_{m-M}=0.4^m$. If galaxies
in the seven selected MK-groups are real members of these groups, than the
observed dispersion of distance moduli should be specified by measurement
errors, i.e. it should be of about $\sim0.4^m$. As one can see from the column
(12) data in the Table~2, the rms meaning $\sigma(m-M)$ over seven groups is
{0.27$^m$, and taking number of galaxies $n_D$ as a weight this value reaches
$\sigma (m-M)=0.40^m$, providing a posteriori an evidence for real membership of
these galaxies.\footnote{The rms variance of distance moduli for four groups
neighbouring to UMa complex and presented in the lower lines of Table~2 is
0.39$^m$; here we eliminate the case of UGC~7774, the galaxy with
$V_{LG}=555$km~s$^{-1}$ but $D$=22.6~Mpc which belongs more likely to Coma~I
cloud than to NGC\,4490 group.}

As it is seen from the last column of Table 2, the infalling zones around
seven groups of the UMa complex overlap each other essentially. Taking also
into account the small scatter of average distances for the groups along the
line of sight ($15.1 - 19.1$~Mpc) one may conclude that all the seven groups
under discussion form a single physical complex (domain).

\section{Concluding remarks}
The latest years provide more and more evidence that besides regular virialized
groups and clusters of galaxies, a population of loose unvirialized structures
exists involving roughly a quarter of all galaxies. The typical dimensions of
such dynamically unrelaxed aggregations (associations, clouds, domains) are
about 30~kpc to 10~Mpc. Population of these structures differs from that one of
regular groups and clusters by low luminosity of galaxies, neutral hydrogen
abundance and active star formation. Tully et al. (2006) even proposed to
distinguish a special category of ``dwarf galaxies associations'', the proximate
example of the latter is the quartet \{NGC 3109, Sex A, Sex B and Antlia\} only
1.3~Mpc away. Makarov \& Uklein (2012) published a list of such dwarf systems
populating the Local supercluster volume. Another example (on larger scales) is
the nearby Canes Venatici~I cloud with dimensions of about 5~Mpc (Karachentsev
et al. 2003). Many of these ``lethargic'' structures are listed in ``Nearby
Galaxies Atlas'' (Tilly \& Fisher, 1987) named ``clouds'' and ``spurs'', i.e.
loose fragments of the large scale structure.

The UMa cloud also represents such unvirialized complex. Among 270 galaxies in
this region 133, i.e. nearly a half, are dwarf systems of morphological types
Ir, Im, BCD ($T=9$ and 10). Most of them appear in the GALEX UV-survey, i.e.
they demonstrate active star formation. The UMa complex has a projected diameter
of about 4~Mpc, roughly the same as its radial dimension (from 15 up to 19 Mpc).

The UMa complex consists of 7 groups having crossing times of (4--7) Gyr, i.e. 2--3
times less than the age of the Universe. Crossing time for the complex itself is
actually equal to the age of the Universe. The total virial mass of the UMa
cloud is $4\cdot10^{13}M_{\odot}$, what is quite typical for a rich group or for
a poor cluster. The mean matter density in UMa, $\Omega_m\simeq 0.08$, appears
to be almost the same as the mean matter density in the Local universe within
the radius of 50~Mpc (Makarov \& Karachentsev, 2011). This is despite the fact
that the Ursa Majoris cloud looks like an overdensity in the Nearby Galaxies Atlas
(Tully and Fisher, 1987). It could be supposed that
UMa domain contains by an order more of dark matter distributed in the volume
between seven groups. However, having the total mass of
$\sim4\cdot10^{14}M_{\odot}$, i.e. comparable with the Virgo cluster mass, the
UMa complex would show the ``Z-wave'' effect of infall. However, this phenomenon
could not be followed in the Hubble diagram. The most natural explanation for
this is that the significant amount of mass in the Universe lies in the empty
space between clusters, where the dark-to-luminous matter ratio is much greater than
100 (Karachentsev, 2012).

According to the bottom panel of
Figure~2, the UMa system of groups is going through nearly free Hubble
expansion. As a whole, the UMa domain moves toward our Local domain with a
peculiar velocity of $(-337\pm28)$~km~s$^{-1}$. Searching for similar objects 
in the Local universe and investigating their kinematics seems to be of a great
importance.

{\bf Acknowledgements.}  This work was partially supported by the Russian
Foundation for Basic Research (grants no. 10-02-00123, 11-02-00639,
RFBR-DFG 12-02-91338), 
by the CNRS, by a grant
from the Space Telescope Science Institute under the NASA contract within
GO12546, and a grant of the Ministry of Education and Science of the Russian
Federation N~14.740.11.0901. O.\,G.\,Nasonova thanks the non-profit Dmitry
Zimin’s Dynasty Foundation for the financial support. We thank Neil Trentham,
the referee, for very useful comments.

\bigskip

{\bf References}

Abazajian K.N., Adelman-McCarthy J.K., Agueros M.A., et al. 2009, ApJS, 182, 543

Bell E.F., McIntosh D.H., Katz N., Weinberg M.D., 2003, ApJS, 149, 289

Jarrett T.N., Chester T., Cutri R. et al. 2000, AJ, 119, 2498

Karachentsev I.D., 2012, Astrophys. Bull., 67, 123

Karachentsev I.D., Nasonova O.G., Courtois H.M., 2011, ApJ, 743, 123

Karachentsev I.D., Sharina M.E., Dolphin A.E., et al. 2003, A \& A, 398, 467

Makarov D.I., Karachentsev I.D., 2011, MNRAS,

Makarov D.I., Uklein R., 2012, Astrophys. Bull., 67,...

Nasonova O.G., de Freitas Pacheco J.A., Karachentsev I.D., 2011, A \& A, 532, 104

Springob C.M., Masters K.L., Haynes M.P. et al. 2009, ApJS, 172, 599

Tonry J.L., Dressler A., Blakeslee J.P., et al. 2001, ApJ, 546, 681

Trentham N., Tully R.B., Verheijen M.A., 2001, MNRAS, 325, 385

Tully R.B., Courtois H.M., 2012, arXiv:1202.3191

Tully R.B., Rizzi L., Shaya E.J., et al. 2009, AJ, 138, 323

Tully R.B., Shaya E.L., Karachentsev I.D., et al. 2008, ApJ, 676, 184

Tully R.B., Rizzi L., Dolphin A.E. et al. 2006, AJ, 132, 729

Tully R.B., Verheijen M.A., Pierce M.J. et al. 1996, AJ, 112, 2471

Tully R.B., 1987, ApJ, 321, 280

Tully R.B., Fisher R.J., 1987, Nearby Galaxies Atlas, Cambridge Univ,
 Cambridge

Tully R.B., Fisher R.J., 1977, A \& A, 54, 661

Wolfinger K., Kilborn V.A., Koribalski B.S., et al. 2012 arXiv:1210.2727

\newpage

\begin{longtable}{llrrrcrr}
\caption{List of 270 galaxies within RA$=[11^h\hspace{-0.4em}.\,0,
13^h\hspace{-0.4em}.\,0]$, DEC$=[+40^{\circ}, +60^{\circ}]$ and
$V_{LG} = [500-1500]$~km~s$^{-1}$}\\
\hhline{========}
 Name          &\multicolumn{1}{c}{RA (2000.0) Dec}&$V_{LG}$ & T  &  K  & Group &  $m-M$  & $D$ Mpc \\
  (1)          &\multicolumn{1}{c}{(2)}            & (3) &(4)& (5)&    (6)  &  (7)  &  (8) \\
\hhline{--------}\endfirsthead
\hhline{========}
  (1)          &\multicolumn{1}{c}{(2)}            & (3) &(4)& (5)&    (6)  &  (7)  &  (8) \\
\hhline{--------}\endhead
\hhline{========}\endfoot
\hhline{========}\endlastfoot
MCG +09-18-066 &110000.2+542532 &1073 &10  &12.6&       &  $-$  &        \\
SDSS J110006.0 &110006.1+541620 &1475 &10  &15.1& N3448 &  $-$  &        \\
UGC06113       &110248.6+520659 &1010 &10  &16.6&       & 31.26 & 17.9   \\
UGC06161       &110649.2+434324 & 773 & 8  &11.3&       & 31.11 & 16.7   \\
UGC06182       &110802.9+533700 &1325 & 8  &10.7&       &  $-$  &        \\
SDSS J110819.9 &110819.9+533628 &1157 &10  &14.8&       &  $-$  &        \\
2MASX J1108394 &110839.5+473154 &1453 & 9  &13.5&       &  $-$  &        \\
UGC06202       &110936.5+505536 & 992 & 7  &13.9&       &  $-$  &        \\
UGC06205       &110958.4+460542 &1417 & 8  &12.4&       & 32.06 & 25.8   \\
SDSS J111100.0 &111100.0+525918 & 877 & 9  &12.0&       &  $-$  &        \\
NGC3556        &111131.0+554027 & 779 & 6  & 7.0&       & 30.30 & 11.5   \\
UGC06249       &111320.7+595433 &1162 & 5  &11.0&       &  $-$  &        \\
UGC06251       &111326.1+533542 & 999 & 9  &12.8& U6251 &  $-$  &        \\
SDSS J111343.6 &111343.6+533848 & 985 &10  &15.4& U6251 &  $-$  &        \\
NGC3600        &111552.0+413528 & 727 & 1  &10.2&       & 30.56 & 12.9   \\
CGCG 268-012   &111700.3+503505 & 900 & 9  &12.9&       &  $-$  &        \\
ARP'S GALAXY   &111934.3+513012 &1388 & 9  &14.5&       &  $-$  &        \\
SDSS           &112017.0+452323 & 707 & 9  &15.1&       &  $-$  &        \\
NGC3631        &112102.9+531011 &1225 & 5  & 8.0& N3631 &  $-$  &        \\
SDSS J112147.5 &112147.6+572048 &1168 & 8  &14.4&       &  $-$  &        \\
SDSS J112235.6 &112235.7+585841 &1354 &10  &14.2&       &  $-$  &        \\
UGC06399       &112323.2+505334 & 864 & 8  &11.1&       & 31.32 & 18.4   \\
NGC3657        &112355.6+525516 &1283 & 0  &10.3& N3631 &  $-$  &        \\
NGC3675        &112608.6+433509 & 789 & 3  & 6.8& N3675 & 30.96 & 15.6   \\
SDSS J112625.9 &112626.0+591738 &1445 &10  &15.1&       &  $-$  &        \\
UGC06446       &112640.5+534448 & 719 & 7  &11.5&       & 31.12 & 16.7   \\
IC0691         &112644.3+590920 &1303 & 9  &10.8&       & 31.76 & 22.5   \\
KDG 078        &112954.5+522413 & 652 &10  &12.5&       &  $-$  &        \\
SDSS J113014.4 &113014.4+595627 &1100 &10  &14.3&       &  $-$  &        \\
NGC3718        &113234.9+530404 &1063 & 1  & 7.8& N3992 & 32.07 & 25.9   \\
SDSS J113237.4 &113237.4+472659 &1497 & 9  &14.8&       &  $-$  &        \\
SDSS J113307.7 &113307.8+472731 &1447 & 9  &13.1&       &  $-$  &        \\
NGC3726        &113321.2+470145 & 904 & 5  & 7.8& N3877 & 30.98 & 15.7   \\
NGC3729        &113349.3+530732 &1097 & 1  & 8.7& N3992 & 31.62 & 21.1   \\
NGC3733        &113501.6+545102 &1267 & 6  &12.1&       & 31.96 & 24.7   \\
MCG +10-17-017 &113518.1+585319 &1134 & 8  &14.0&       &  $-$  &        \\
UGC06566       &113543.6+581133 &1332 & 8  &14.1& N3838 &  $-$  &        \\
UGC06575       &113626.5+581129 &1316 & 6  &11.4& N3838 & 32.79 & 36.1   \\
NGC3756        &113648.0+541737 &1367 & 4  & 8.8& N3756 & 31.22 & 17.5   \\
NGC3757        &113702.9+582456 &1368 &$-$5  & 9.6& N3838 &  $-$  &        \\
NGC3769        &113744.1+475335 & 780 & 3  & 9.2& N3769 & 30.98 & 15.7   \\
2MASX J1137444 &113744.4+540245 & 986 & 9  &13.8& N3992 &  $-$  &        \\
NGC3769A       &113751.4+475253 & 834 & 9  &13.5& N3769 &  $-$  &        \\
UGC06604       &113808.6+584530 &1426 &$-$5  &10.2& N3838 &  $-$  &        \\
MRK 1450       &113835.6+575227 &1087 & 9  &14.3&       &  $-$  &        \\
UGC06611       &113851.5+430952 &1164 & 7  &13.1&       &  $-$  &        \\
NGC3782        &113920.7+463048 & 780 & 7  &10.7& N3769 & 30.82 & 14.6   \\
NGC3795A       &113921.3+581607 &1264 & 6  &10.6& N3838 & 31.88 & 23.8   \\
SDSS J113924.7 &113924.8+413558 &1137 & 8  &15.8&       &  $-$  &        \\
SDSS J113930.2 &113930.3+432428 & 867 &10  &15.5& N3675 &  $-$  &        \\
SDSS           &113948.4+543116 & 805 & 9  &14.2&       &  $-$  &        \\
SDSS J113948.7 &113948.8+463711 & 744 &10  &15.1& N3769 &  $-$  &        \\
PGC166114      &114003.3+462851 & 777 &10  &15.7& N3769 &  $-$  &        \\
UGC06628       &114006.7+455634 & 888 & 9  &10.7& N3877 &  $-$  &        \\
NGC3795        &114006.8+583647 &1271 & 4  &10.6& N3838 & 32.25 & 28.2   \\
SDSS J114033.0 &114033.0+573335 &1116 &$-$5  &13.7&       &  $-$  &        \\
SDSS J114035.6 &114035.6+460728 & 882 &10  &14.0& N3877 &  $-$  &        \\
NGC3794        &114053.4+561207 &1472 & 6  &11.3&       & 31.20 & 17.4   \\
SDSS J114106.7 &114106.8+534752 &1349 & 8  &14.4& N3756 &  $-$  &        \\
CGCG 242-075   &114122.0+462336 & 856 & 9  &13.9& N3769 &  $-$  &        \\
UGC06667       &114226.3+513553 &1042 & 6  &11.7& N3992 & 31.18 & 17.2   \\
SBS 1139+550   &114227.2+544908 &1368 & 0  &12.1& N3756 &  $-$  &        \\
UGC06682       &114309.1+590621 &1431 & 8  &12.4& N3838 &  $-$  &        \\
SDSS           &114330.7+531113 &1371 & 9  &14.6& N3756 &  $-$  &        \\
UGC06685       &114331.1+552844 &1091 & 6  &12.9& N3992 &  $-$  &        \\
NGC3838        &114413.8+575654 &1420 &$-$1  & 9.3& N3838 &  $-$  &        \\
UGC06713       &114425.0+485007 & 955 & 8  &11.5& N3877 &  $-$  &        \\
CGCG 292-024   &114452.1+575225 &1365 & 9  &12.7& N3838 &  $-$  &        \\
SDSS J114525.7 &114525.7+482907 & 932 &10  &15.1& N3877 &  $-$  &        \\
NGC3850        &114535.6+555313 &1242 & 5  &11.9& N3992 & 31.00 & 15.8   \\
NGC3870        &114556.6+501159 & 817 & 7  &10.8&       & 30.21 & 11.0   \\
SDSS J114604.5 &114604.5+563356 &1114 &10  &15.6& N3992 &  $-$  &        \\
NGC3877        &114607.8+472941 & 950 & 5  & 7.7& N3877 & 30.95 & 15.5   \\
SDSS J114613.4 &114613.4+541034 &1125 &10  &15.1& N3992 &  $-$  &        \\
SDSS J114628.2 &114628.3+532444 & 988 &10  &15.3& N3992 &  $-$  &        \\
SDSS J114634.0 &114634.1+554917 &1159 &10  &14.5& N3992 &  $-$  &        \\
SDSS J114643.2 &114643.3+571358 &1118 &10  &15.1& N3992 &  $-$  &        \\
SDSS J114702.8 &114702.9+541717 &1458 &10  &16.2&       &  $-$  &        \\
ARK 324        &114745.2+595311 &1343 & 9  &12.8& N4036 &  $-$  &        \\
SDSS J114751.3 &114751.4+535048 &1097 & 9  &13.1& N3992 &  $-$  &        \\
SDSS J114754.7 &114754.7+582151 &1434 &10  &14.6& N3838 &  $-$  &        \\
UGC06773       &114800.5+494830 & 985 & 8  &11.2& N3992 & 30.78 & 14.3   \\
SDSS J114820.2 &114820.2+562046 &1123 &10  &14.5& N3992 &  $-$  &        \\
SDSS J114829.3 &114829.3+570755 &1396 &10  &15.4& N3838 &  $-$  &        \\
UGC06776       &114835.8+434320 & 762 & 8  &12.8& N4111 &  $-$  &        \\
NGC3893        &114838.2+484239 &1025 & 5  & 7.9& N3877 & 31.14 & 16.9   \\
SDSS           &114845.2+492130 & 785 & 9  &14.9&       &  $-$  &        \\
SDSS J114855.4 &114855.5+473458 & 994 &10  &13.9& N3877 &  $-$  &        \\
NGC3896        &114856.4+484029 & 961 & 9  &11.6& N3877 &  $-$  &        \\
$[$HS98$]$ 219 &114900.1+572353 &1299 & 9  &14.0& N3992 &  $-$  &        \\
NGC3898        &114915.4+560504 &1266 & 2  & 7.7& N3992 &  $-$  &        \\
SDSS J114929.6 &114929.7+560155 &1010 &10  &15.7& N3992 &  $-$  &        \\
SDSS J114930.9 &114930.9+442433 & 870 &10  &15.2& N4111 &  $-$  &        \\
NGC3906        &114940.5+482534 &1013 & 6  &11.0& N3877 &  $-$  &        \\
SBS 1147+520   &114954.5+514411 &1034 & 9  &16.3& N3992 &  $-$  &        \\
KKH73          &115006.4+554700 & 685 &10  &14.9&       &  $-$  &        \\
UGC06802       &115006.7+515117 &1322 & 6  &12.4& N3992 & 31.67 & 21.6   \\
UGC06805       &115012.3+420428 &1055 & 9  &11.5&       &  $-$  &        \\
NGC3913        &115038.9+552114 &1045 & 6  &10.8& N3992 &  $-$  &        \\
NGC3917        &115045.5+514927 &1039 & 5  & 8.8& N3992 & 31.04 & 16.1   \\
UGC06818       &115046.5+454824 & 855 & 7  &11.7& N4111 & 31.63 & 21.2   \\
UGC06816       &115047.7+562721 & 984 & 9  &11.9& N3992 & 32.16 & 27.0   \\
MRK 1460       &115050.0+481505 & 843 & 9  &14.5& N3877 &  $-$  &        \\
SDSS J115056.1 &115056.1+483154 &1023 &10  &14.1& N3877 &  $-$  &        \\
SDSS J115059.6 &115059.6+475750 & 988 &10  &15.1& N3877 &  $-$  &        \\
NGC3922        &115113.4+500925 &1016 & 0  &10.0& N3992 &  $-$  &        \\
NGC3931        &115113.4+520003 &1001 &$-$3  &10.6& N3992 &  $-$  &        \\
SDSS J115126.7 &115126.8+494734 &1270 &10  &14.8&       &  $-$  &        \\
NGC3928        &115147.6+484059 &1043 &$-$1  & 9.7& N3877 & 30.97 & 15.6   \\
SDSS J115153.6 &115153.7+530558 &1154 &10  &14.3& N3992 &  $-$  &        \\
UGC06840       &115207.0+520629 &1093 & 8  &11.8& N3992 & 31.02 & 16.0   \\
SDSS J115233.4 &115233.4+481735 &1126 &10  &14.7&       &  $-$  &        \\
UGC06849       &115239.2+500216 &1091 & 8  &11.9& N3992 &  $-$  &        \\
NGC3938        &115249.5+440715 & 843 & 5  & 7.8& N4111 & 31.27 & 17.9   \\
SDSS J115332.9 &115333.0+455422 & 848 &10  &14.3& N4111 &  $-$  &        \\
NGC3949        &115341.4+475132 & 854 & 4  & 8.6& N3877 & 31.19 & 17.3   \\
NGC3953        &115348.9+521936 &1126 & 4  & 7.0& N3992 & 31.22 & 17.5   \\
SDSS           &115352.3+512938 & 568 & 9  &15.1&       &  $-$  &        \\
SDSS J115356.9 &115357.0+551017 &1338 &$-$1  &13.3& N3992 &  $-$  &        \\
SDSS J115441.2 &115441.2+463636 &1094 & 9  &14.8&       &  $-$  &        \\
SDSS J115457.9 &115458.0+443335 & 880 & 9  &13.8& N4111 &  $-$  &        \\
SDSS           &115506.0+440612 & 664 &10  &15.9& N4111 &  $-$  &        \\
SDSS           &115513.1+441308 & 653 & 9  &14.6& N4111 &  $-$  &        \\
KDG 081        &115514.3+440902 & 766 &10  &13.7& N4111 &  $-$  &        \\
UGC06894       &115524.4+543926 & 945 & 7  &13.6& N3992 & 31.57 & 20.6   \\
SBS 1153+565   &115537.1+561511 &1062 & 9  &13.2& N3992 &  $-$  &        \\
NGC3972        &115545.1+551915 & 933 & 4  & 9.6& N3992 & 31.34 & 18.5   \\
SDSS J115551.8 &115551.8+450946 &1033 &10  &14.3&       &  $-$  &        \\
SDSS J115603.7 &115603.7+522618 & 946 &10  &14.7& N3992 &  $-$  &        \\
UGC06912       &115614.4+581149 &1457 & 8  &12.2&       & 31.16 & 17.1   \\
NGC3982        &115628.1+550731 &1198 & 3  & 8.8& N3992 & 31.70 & 21.9   \\
UGC06917       &115628.8+502542 & 979 & 7  &11.2& N3992 & 31.33 & 18.4   \\
UGC06919       &115637.5+553800 &1375 & 4  &11.6&       &  $-$  &        \\
NGC3985        &115642.1+482002 &1002 & 8  &10.3& N3877 & 31.20 & 17.4   \\
SDSS J115644.3 &115644.3+490118 &1048 &10  &15.4& N3877 &  $-$  &        \\
SDSS J115647.7 &115647.8+585820 &1333 &10  &14.4& N4036 &  $-$  &        \\
UGC06923       &115649.4+530937 &1144 & 8  &11.3& N3992 & 31.09 & 16.5   \\
UGC06922       &115652.1+504901 & 960 & 4  &11.9& N3992 &  $-$  &        \\
UGC06926       &115655.4+573047 &1182 & 8  &13.0& N3992 &  $-$  &        \\
SDSS J115701.8 &115701.9+552511 &1306 & 9  &12.3& N3992 &  $-$  &        \\
SDSS           &115703.1+553512 & 855 & 9  &14.3& N3992 &  $-$  &        \\
UGC06930       &115717.4+491659 & 839 & 6  &11.2& N4157 &  $-$  &        \\
UGC06931       &115724.9+575548 &1289 & 8  &11.8&       & 31.43 & 19.3   \\
NGC3990        &115735.6+552731 & 788 &$-$2  & 9.5& N3992 & 30.06 & 10.3   \\
NGC3992        &115736.0+532228 &1129 & 4  & 6.9& N3992 & 31.80 & 22.9   \\
UGC06940       &115747.6+531404 &1192 & 4  &14.0& N3992 & 31.98 & 24.9   \\
NGC3998        &115756.1+552713 &1145 &$-$2  & 7.4& N3992 & 30.75 & 14.1   \\
SDSS           &115802.2+512057 & 632 & 9  &14.1&       &  $-$  &        \\
2MASX J1158109 &115811.0+580923 &1064 & 9  &13.7&       &  $-$  &        \\
MCG +08-22-048 &115811.6+485253 & 896 &10  &12.2& N3877 &  $-$  &        \\
SDSS J115813.6 &115813.7+552317 &1058 &$-$1  &12.5& N3992 &  $-$  &        \\
UGC06956       &115825.6+505501 & 987 & 8  &12.1& N3992 &  $-$  &        \\
NGC4013        &115831.4+435648 & 875 & 3  & 7.6& N4111 & 31.38 & 18.9   \\
IC0749         &115834.0+424402 & 836 & 6  &10.4& N4111 &  $-$  &        \\
SDSS J115834.3 &115834.3+532044 &1232 &$-$1  &13.1& N3992 &  $-$  &        \\
NGC4010        &115837.9+471541 & 957 & 7  & 9.6& N3877 & 31.16 & 17.1   \\
UGC06969       &115847.6+532529 &1195 & 7  &12.6& N3992 & 31.72 & 22.1   \\
SDSS J115849.1 &115849.2+551825 &1030 &$-$1  &13.1& N3992 &  $-$  &        \\
SDSS J115849.7 &115849.7+462753 & 881 &10  &14.6& N3877 &  $-$  &        \\
IC0750         &115852.2+424321 & 730 & 2  & 8.1& N4111 & 31.85 & 23.4   \\
UGCA 259       &115853.3+454404 &1198 &10  &13.8&       & 31.27 & 17.9   \\
CGCG215-13     &115856.8+441134 & 733 & 8  &12.8& N4111 &  $-$  &        \\
UGC06983       &115909.3+524227 &1157 & 6  &10.5& N3992 & 31.51 & 20.0   \\
SDSS J115921.8 &115921.8+564646 &1181 & 9  &12.7& N3992 &       &        \\
NGC4026        &115925.2+505742 &1032 &$-$2  & 7.6& N3992 & 30.67 & 13.6   \\
SDSS J115925.3 &115925.3+570409 & 944 &10  &14.7&       &  $-$  &        \\
KUG1156+42     &115929.1+422057 & 815 &10  &14.5& N4111 &  $-$  &        \\
SDSS           &115937.0+425716 & 695 &10  &14.9& N4111 &  $-$  &        \\
SDSS J115943.2 &115943.3+533639 &1066 &10  &14.5& N3992 &  $-$  &        \\
SDSS J115950.8 &115950.8+502955 & 974 &10  &13.7& N3992 &  $-$  &        \\
UGC06988       &115951.7+553955 & 814 & 8  &13.0&       &  $-$  &        \\
2MASX J1159562 &115956.2+532945 &1122 & 9  &13.9& N3992 &  $-$  &        \\
MCG +08-22-051 &115957.7+493350 &1196 & 9  &13.7& N3992 &  $-$  &        \\
PGC166118      &115958.6+444306 &1210 &10  &14.2&       &  $-$  &        \\
SDSS J120002.4 &120002.4+424723 &1031 &$-$1  &12.9& N4111 &  $-$  &        \\
UGC06992       &120018.9+503910 & 820 & 7  &11.4& N4157 &  $-$  &        \\
UGCA 262       &120035.4+474626 & 624 &10  &14.4& N4258 &  $-$  &        \\
MCG +09-20-060 &120044.4+543315 &1361 &10  &13.4& N3992 &  $-$  &        \\
MCG +09-20-063 &120100.3+550133 &1198 & 9  &13.5& N3992 &  $-$  &        \\
UGC06999       &120101.2+495446 & 995 &10  &13.9& N3992 &  $-$  &        \\
SDSS J120139.6 &120139.6+551231 &1286 & 9  &12.4& N3992 &  $-$  &        \\
2MASX J1201501 &120150.1+550842 &1199 & 9  &13.9& N3992 &  $-$  &        \\
SDSS J120204.3 &120204.3+563649 &1311 & 9  &15.4& N3992 &  $-$  &        \\
UGC07022       &120243.7+451128 & 728 & 8  &12.8& N4111 &  $-$  &        \\
SDSS J120255.5 &120255.5+554906 &1138 &10  &14.1& N3992 &  $-$  &        \\
SDSS J120259.9 &120300.0+473915 & 655 &10  &14.6& N4258 &  $-$  &        \\
NGC4051        &120309.6+443153 & 740 & 4  & 7.7& N4111 & 30.80 & 14.5   \\
2MASX J1203230 &120322.9+434439 &1090 & 9  &14.2& N4111 &  $-$  &        \\
SDSS J120330.7 &120330.7+550306 &1222 &10  &15.8& N3992 &  $-$  &        \\
NGC4085        &120522.7+502110 & 820 & 5  & 9.1& N4157 & 31.37 & 18.8   \\
NGC4088        &120534.2+503220 & 829 & 5  & 7.5& N4157 & 31.04 & 16.1   \\
SDSS J120549.5 &120549.5+504729 & 889 &10  &14.5& N4157 &  $-$  &        \\
UGC07089       &120558.1+430843 & 808 & 7  &11.1& N4111 & 30.68 & 13.7   \\
SDSS J120559.6 &120559.6+425409 & 788 &10  &15.0& N4111 &  $-$  &        \\
NGC4096        &120601.1+472842 & 625 & 5  & 7.8& N4258 & 30.60 & 13.2   \\
NGC4100        &120608.1+493459 &1142 & 4  & 8.0& N3992 & 31.47 & 19.7   \\
UGC07094       &120610.8+425721 & 812 & 8  &12.5& N4111 & 30.57 & 13.0   \\
NGC4102        &120623.1+524239 & 923 & 3  & 7.7& N3992 & 31.41 & 19.1   \\
SDSS           &120625.4+422605 &1000 & 8  &14.8& N4111 &  $-$  &        \\
SDSS J120637.9 &120637.9+544558 & 936 &10  &19.9& N3992 &  $-$  &        \\
NGC4111        &120703.1+430355 & 818 &$-$1  & 7.5& N4111 & 30.88 & 15.0   \\
NGC4117        &120746.1+430735 & 969 & 0  &10.0& N4111 &  $-$  &        \\
SDSS J120751.6 &120751.6+413347 &1098 & 9  &15.0& N4111 &  $-$  &        \\
NGC4118        &120752.9+430640 & 677 & 9  &13.2& N4258 &  $-$  &        \\
SDSS J120810.7 &120810.7+554447 &1179 & 9  &14.2& N3992 &  $-$  &        \\
SDSS J120824.5 &120824.5+412405 & 955 &10  &14.5& N4111 &  $-$  &        \\
SDSS           &120847.8+511147 & 646 & 9  &14.3& N4258 &  $-$  &        \\
UGC07129       &120855.1+414427 & 950 & 2  &10.6& N4111 & 31.77 & 22.6   \\
NGC4138        &120929.8+434107 & 926 & 1  & 8.2& N4111 & 30.70 & 13.8   \\
NGC4142        &120930.2+530618 &1238 & 6  &10.8& N3992 & 31.93 & 24.3   \\
SDSS J120931.7 &120931.8+545618 &1061 & 9  &14.0& N3992 &  $-$  &        \\
NGC4143        &120936.1+423203 &1002 &$-$2  & 7.8& N4111 & 31.00 & 15.8   \\
UGC07146       &120949.1+431405 &1101 & 8  &13.5& N4111 & 31.04 & 16.1   \\
UGC07176       &121055.9+501718 & 959 & 8  &14.6& N4157 &  $-$  &        \\
SBS 1208+531   &121100.7+524957 & 970 & 9  &14.7& N3992 &  $-$  &        \\
NGC4157        &121104.4+502905 & 842 & 4  & 7.4& N4157 & 31.23 & 17.6   \\
BTS97          &121122.6+501611 & 829 & 9  &12.6& N4157 &  $-$  &        \\
SDSS           &121135.0+473927 & 805 &10  &15.2& N4157 &  $-$  &        \\
MCG +08-22-083 &121155.7+465854 &1022 & 8  &13.5& N4217 &  $-$  &        \\
SDSS J121255.1 &121255.2+440527 &1019 & 9  &15.6& N4111 &  $-$  &        \\
SBS1210+53     &121255.9+532738 &1042 & 9  &13.8& N3992 &  $-$  &        \\
UGC07218       &121256.5+521555 & 862 & 8  &12.3& N3992 & 30.88 & 15.0   \\
SDSS J121304.9 &121304.9+530620 &1374 & 9  &14.0&       &  $-$  &        \\
NGC4183        &121316.9+434155 & 970 & 6  & 9.8& N4111 & 31.01 & 15.9   \\
SBS 1211+540   &121402.5+534517 & 996 & 9  &14.7& N3992 &  $-$  &        \\
UGC07267       &121523.6+512100 & 550 & 8  &11.5& N4258 & 30.55 & 12.9   \\
UGC07271       &121533.4+432603 & 585 & 7  &13.2& N4258 & 30.38 & 11.9   \\
SDSS J121537.1 &121537.1+441710 & 921 &10  &14.2& N4111 &  $-$  &        \\
NGC4218        &121546.4+480751 & 787 & 7  &10.9& N4346 & 31.19 & 17.3   \\
NGC4217        &121550.9+470530 &1085 & 3  & 7.6& N4217 & 31.45 & 19.5   \\
SDSS           &121551.6+473017 & 702 & 9  &14.7& N4346 &  $-$  &        \\
MCG +08-22-086 &121602.2+464358 &1112 & 7  &14.1& N4217 &  $-$  &        \\
NGC4220        &121611.7+475300 & 987 & 1  & 8.1& N4217 & 31.37 & 18.8   \\
UGC07301       &121642.1+460444 & 759 & 7  &12.7& N4346 & 31.59 & 20.8   \\
UGC07320       &121728.6+444840 & 568 &10  &13.0& N4258 & 29.82 &  9.2   \\
NGC4242        &121730.2+453710 & 567 & 7  & 8.9& N4258 & 29.43 &  7.7   \\
2MASX          &121731.9+475942 & 756 & 9  &13.2& N4346 &  $-$  &        \\
NGC4248        &121749.9+472433 & 551 & 8  &10.6& N4258 & 29.35 &  7.4   \\
SDSS J121811.0 &121811.0+465501 & 530 & 8  &14.5& N4258 &  $-$  &        \\
SDSS J121840.1 &121840.1+455435 &1106 & 9  &12.7& N4217 &  $-$  &        \\
NGC4258        &121857.5+471814 & 509 & 4  & 5.5& N4258 & 29.47 &  7.8   \\
SDSS J121915.1 &121915.1+444802 & 961 & 9  &15.0& N4111 &  $-$  &        \\
KK133          &121932.8+432311 & 601 &10  &15.3&       &  $-$  &        \\
UGC07391       &122016.2+455430 & 665 & 8  &13.7&       & 30.36 & 11.8   \\
UGC07392       &122017.5+480816 & 861 & 8  &13.5& N4346 &  $-$  &        \\
NGC4288        &122038.1+461730 & 590 & 7  &10.4& N4258 &  $-$  &        \\
UGC07401       &122048.4+474933 & 804 &10  &13.5& N4346 & 31.05 & 16.2   \\
UGC07408       &122115.0+454841 & 515 & 9  &11.1& N4258 &  $-$  &        \\
SDSS J122308.0 &122308.1+530120 & 973 & 9  &15.1&       &  $-$  &        \\
NGC4346        &122327.9+465938 & 822 &$-$2  & 8.2& N4346 & 30.78 & 14.3   \\
NGC4389        &122535.1+454105 & 772 & 4  & 9.1& N4346 & 30.73 & 14.0   \\
UGC07534       &122608.1+581921 & 838 & 8  &12.1&       & 30.11 & 10.5   \\
UGC07608       &122844.2+431327 & 580 & 8  &11.2& N4490 & 29.74 &  8.9   \\
NGC4460        &122845.6+445151 & 542 & 1  & 9.1& N4258 & 29.91 &  9.6   \\
NGC4485        &123031.1+414204 & 517 & 8  &10.6& N4490 &  $-$  &        \\
NGC4490        &123036.4+413837 & 622 & 7  & 7.3& N4490 & 28.82 &  5.8   \\
SDSS J123106.0 &123106.1+444449 &1000 & 9  &13.2&       &  $-$  &        \\
MAPS-NGP O-218 &123109.0+420539 & 602 &10  &15.4& N4490 &  $-$  &        \\
UGC07690       &123226.9+424215 & 581 & 8  &12.2& N4490 & 30.25 & 11.2   \\
UGC07751       &123511.8+410339 & 641 &10  &14.0& N4490 & 29.50 &  7.9   \\
UGC07774       &123622.5+400019 & 556 & 7  &12.5& N4490 & 31.77 & 22.6   \\
UGC07827       &123938.9+444914 & 609 &10  &13.5& N4258 & 29.43 &  7.7   \\
NGC4618        &124132.9+410903 & 585 & 6  & 8.7& N4490 & 29.49 &  7.9   \\
NGC4625        &124152.7+411626 & 651 & 7  & 9.7& N4490 & 29.49 &  7.9   \\
UGC07903       &124345.0+535732 & 546 &10  &14.8&       &  $-$  &        \\
UGCA 297       &124623.3+481407 & 977 & 9  &13.1&       &  $-$  &        \\
UGC07950       &124656.5+513647 & 592 & 9  &12.1&       & 29.75 &  8.9   \\
NGC4707        &124822.9+510953 & 557 &10  &11.7&       & 28.82 &  5.8   \\
SDSS J124931.0 &124931.0+442133 & 575 & 9  &14.8&       &  $-$  &        \\
NGC4800        &125437.8+463152 & 952 & 2  & 8.3&       &  $-$  &        \\
\end{longtable}

\newpage

\hspace{-2.3cm}%
\vbox{\tabcolsep=0.8ex\begin{longtable}{lcrrrrrrrrrrrrr}
\caption{Average parameters of the UMa groups}\\
\hhline{===============}
Group &  RA,DEC &$<V_{LG}>$& $n_v$& T& $\sigma_v$& $R_h$& $\lg L_k$& $\lg M_v$ &$M_v/L_k$& $<m-M>$ & $\sigma_m$  & D  & $n_D$& $R_0$ \\
\hhline{---------------}\endfirsthead
\hhline{===============}
Group &  RA,DEC &$<V_{LG}>$& $n_v$& T& $\sigma_v$& $R_h$& $\lg L_k$& $\lg M_v$ &$M_v/L_k$& $<m-M>$ & $\sigma_m$  & D  & $n_D$& $R_0$ \\
\hhline{---------------}\endhead
N3769 &1137+4753&  780 &  6 & 3&  42&  35&  9.81&  11.99& 151 & 30.90 & 0.08 &15.1&  2 &0.76 \\
      &         &      &    &  &  25&  49& 10.11&  11.69&  38 &       &      &    &    &0.60 \\
      &         &      &    &  &    &    &      &       &     &       &      &    &    &     \\
N3877 &1146+4729&  955 & 21 & 5&  65& 239& 11.05&  12.57&  33 & 31.08 & 0.10 &16.4&  7 &1.18 \\
      &         &      &    &  &  61& 299& 11.25&  12.61&  23 &       &      &    &    &1.22 \\
      &         &      &    &  &    &    &      &       &     &       &      &    &    &     \\
N3992 &1157+5322& 1093 & 74 & 4& 122& 452& 11.68&  13.34&  46 & 31.34 & 0.49 &18.5& 26 &2.12 \\
      &         &      &    &  & 119& 556& 11.86&  13.41&  37 &       &      &    &    &2.24 \\
      &         &      &    &  &    &    &      &       &     &       &      &    &    &     \\
N4111 &1207+4304&  829 & 35 &$-$1& 106& 212& 11.14&  12.80&  45 & 31.13 & 0.40 &16.8& 13 &1.30 \\
      &         &      &    &  & 103& 306& 11.46&  12.93&  29 &       &      &    &    &1.43 \\
      &         &      &    &  &    &    &      &       &     &       &      &    &    &     \\
N4157 &1211+5029&  831 & 10 & 3&  59& 150& 10.82&  12.13&  21 & 31.21 & 0.08 &17.5&  3 &0.90 \\
      &         &      &    &  &  51& 230& 11.19 & 12.21&  11 &       &      &    &    &0.95 \\
      &         &      &    &  &    &    &      &       &     &       &      &    &    &     \\
N4217 &1215+4705& 1085 &  5 & 3&  55& 224& 10.83&  12.20&  23 & 31.41 & 0.04 &19.1&  2 &0.89 \\
      &         &      &    &  &  55& 288& 11.05&  12.31&  18 &       &      &    &    &0.97 \\
      &         &      &    &  &    &    &      &       &     &       &      &    &    &     \\
N4346 &1223+4700&  782 &  8 &$-$2&  45& 286& 10.29&  11.92&  42 & 31.07 & 0.31 &16.4&  5 &0.49 \\
      &         &      &    &  &  42& 435& 10.65&  11.74&  12 &       &      &    &    &0.50\\
      &         &      &    &  &    &    &      &       &     &       &      &    &    &     \\
\hhline{---------------}
N3838 &1144+5757& 1368 & 11 & 0&  63& 202& 10.60&  12.19&  39 & 32.31 & 0.37 &29.0&  3 &0.89 \\
      &         &      &    &  &  49& 313& 10.98&  12.16&  15 &       &      &    &    &0.86 \\
      &         &      &    &  &    &    &      &       &     &       &      &    &    &     \\
N4151 &1210+3924& 1031 &  8 & 2&  69& 348& 11.03&  12.56&  34 & 30.83 & 0.21 &14.7&  6 &1.17  \\
      &         &      &    &  &  66& 362& 11.07&  12.54&  30 &       &      &    &    &1.16  \\
      &         &      &    &  &    &    &      &       &     &       &      &    &    &       \\
N4258 &1219+4718&  567 & 18 & 4&  81& 254& 10.97&  12.46&  31 & 29.88 & 0.48 & 9.5&  9 &1.08  \\
      &         &      &    &  &  78& 320& 11.17&  12.53&  23 &       &      &    &    &1.14  \\
      &         &      &    &  &    &    &      &       &     &       &      &    &    &      \\
N4490 &1230+4138&  583 &  8 & 7&  45&  98& 10.36&  11.84&  30 & 29.55 & 0.40 & 8.1&  6 &0.68  \\
      &         &      &    &  &  44&  98& 10.36&  11.82&  29 &       &      &    &    &0.67  \\
\hhline{===============}
\end{longtable}}

\end{document}